\documentclass[preprint]{aastex}
\usepackage{graphicx}
\usepackage{latexsym}

\newcommand{\HI}{\protect\ion{H}{1}}
\newcommand{\mhi}{$M_{HI}$}
\newcommand{\msun}{$M_\odot$}
\newcommand{\cmsq}{cm$^{-2}$}
\newcommand{\nhi}{$N_{HI}$}
\newcommand{\kms}{~km\,s$^{-1}$}

\begin{document}


\title{Green Bank Telescope observations of low column density \HI\ around NGC~2997 and NGC~6946}
\author{D.J. Pisano\altaffilmark{1}}
\affil{Department of Physics \& Astronomy, West Virginia University, P.O. Box 6315, Morgantown, WV, 26506, USA}
\altaffiltext{1}{Adjunct Assistant Astronomer at National Radio Astronomy Observatory, 
P.O. Box 2, Green Bank, WV 24944}
\email{djpisano@mail.wvu.edu}

\begin{abstract}
Observations of ongoing \HI\ accretion in nearby galaxies have only identified about 10\% of the needed fuel to sustain star formation in these galaxies.  Most of these
observations have been conducted using interferometers and may have missed lower column density, diffuse, \HI\ gas that may trace the missing 90\% of gas.  Such
gas may represent the so-called ``cold flows" predicted by current theories of galaxy formation to have never been heated above the virial temperature of the dark matter
halo.   As a first attempt to identify such cold flows around nearby galaxies and complete the census of \HI\ down to \nhi$\sim$10$^{18}$\cmsq, I used the Robert C. Byrd Green Bank Telescope 
(GBT) to map the circumgalactic (r$\lesssim$100-200 kpc) \HI\ environment around NGC~2997 and NGC~6946.  The resulting GBT observations cover a four square degree area around each galaxy with a 
5$\sigma$ detection limit of \nhi$\sim$10$^{18}$\cmsq\ over a 20~\kms\ linewidth.  This project complements absorption line studies, which are well-suited to the regime of lower \nhi.  
Around NGC~2997, the GBT \HI\ data reveal an extended \HI\ disk and all of its surrounding gas-rich satellite galaxies, 
but no filamentary features.  Furthermore, the \HI\ mass as measured with the GBT is only 7\% higher than past interferometric measurements.  After correcting for resolution differences, the \HI\ extent of the galaxy 
is 23\% larger at the \nhi$=$1.2$\times$10$^{18}$\cmsq\ level as measured by the GBT.  On the other hand, the \HI\ observations of NGC~6946 reveal a filamentary feature apparently connecting NGC~6946
with its nearest companions.  This \HI\ filament has \nhi$\sim$5$\times$10$^{18}$\cmsq\ and a FWHM of 55$\pm$5 \kms and was invisible in past interferometer observations.  The properties 
of this filament are broadly consistent with being a cold flow or debris from a past tidal interaction between NGC~6946 and its satellites.  
\end{abstract}


\section{Introduction}

There have been many detections of low mass neutral hydrogen (\HI) clouds accreting onto nearby galaxies \citep[see][and references therein]{sancisi08}.  The inferred accretion rate of
such clouds, however, is only $\gtrsim$0.1-0.2 \msun yr$^{-1}$; an order of magnitude lower that what is needed to fuel continued star formation in galaxies \citep{sancisi08, kauffmann10}.
This discrepancy either implies that star formation in galaxies will cease in the next few billion years or that our census of gas around galaxies in incomplete.  It should come as no 
surprise that \HI\ observations may be missing a large reservoir of gas around galaxies.  Most past detections of \HI\ clouds have been made with interferometers with \HI\ column density
sensitivities of $\sim$10$^{19}$\cmsq.  Below this level hydrogen is believed to be mostly ionized and so only a small fraction of the gas will be visible as \HI\ \citep{maloney93}. Such
gas is likely to be more diffuse than the \HI\ clouds visible at higher column densities, and since interferometer surveys are blind to gas distributed over large angular scales, the current census will be deficient.    

There are very few observations of \HI\ emission below \nhi$\sim$10$^{19}$\cmsq.  The most prominent survey was that of \citet{braun04}, who identified a low column density, 
diffuse, \HI\ filament connecting M~31 and M~33.  They attributed this filament to the cosmic web similar to those seen in simulations \citep{popping09}.  
In this case, the filament would be an example of a ``cold flow" as predicted by \citet{birnboim03} and \citet{keres05,keres09}.  Cold flows should be the dominant form of 
accretion for galaxies with M$_{halo}\lesssim$10$^{11.4}$\msun\ and in the lowest density environments, n$_{gal}\lesssim$1 h$^3$Mpc$^{-3}$ \citep{keres05}.  M~31 
has M$_{dyn}\sim$1.3$\times$10$^{12}$\msun\ \citep{corbelli10}, so cold accretion is an unlikely explanation.  On the other hand, numerous other authors have suggested that this \HI\ 
filament can be explained as a tidal feature from a past encounter between the two galaxies \citep{bekki08,putman09}.  This hypothesis is supported 
by the extensive stellar streams seen around M~31 and M~33 \citep{ibata01,ibata07,ferguson02} as well as apparent tidal features seen in the \HI\ distribution 
of M~33 \citep{putman09} and simulations that broadly match the distribution of material in the system \citep{bekki08}.  While more sensitive, higher resolution observations by \citet{wolfe13}, 
show that this \HI\ filament is actually composed of small \HI\ clouds with \mhi$\sim$10$^{4-5}$\msun, diameters less than a few kpc, and \nhi$\lesssim$10$^{18}$\cmsq, the best way to distinguish 
between the two possible origins of the diffuse gas between M~31 and M~33 is to identify other instances of similar gaseous features.  Despite the clumpy nature of this filament, if it were located around
a more distant galaxy, it would appear as a continuous structure.  Furthermore, if the diffuse \HI\ filament is part of a cold flow, then such features should be seen around other galaxies with similar properties.  
If the filament is tidal, then analogs should only be seen around galaxies 
that have recently undergone an interaction.  As a pilot study to search for analogous \HI\ filaments and to begin to complete the census of \HI\ emission below 
\nhi$\sim$10$^{19}$\cmsq, I chose two galaxies with similar properties to M~31:  NGC~2997 and NGC~6946.

NGC~2997 is a relatively nearby, D$\sim$12 Mpc, late-type spiral galaxy (Sc) that resides in the loose group LGG~180, 
which is composed of 8 gas-rich galaxies \citep{hess09,pisano07,pisano11}.  It has a measured \HI\ 
rotation velocity of 226 km/s, corresponding to M$_{dyn}=$2.1$\times$10$^{11}$\msun\ \citep{hess09}, and an absolute 
magnitude of -20.7 mag.  It is nearly identical to M~31 in luminosity, but has a lower mass and its nearest known 
companion is $\sim$100 kpc away.  NGC~2997 also has a higher star formation 
rate, 5 M$_\odot$/yr, than M~31, $\sim$1 \msun\ yr$^{-1}$ \citep{williams03}.  
\citet{pisano07,pisano11} observed a $\sim$1 Mpc$^2$ area of the group 
LGG~180 in \HI\ using the Parkes radio telescope down to a rms mass sensitivity
of 1$\times$10$^6$\msun\ and a column density sensitivity, for emission filling the 14$\arcmin$ beam, of 
3.4$\times$10$^{16}$\cmsq\ per 3.3 \kms\ channel and found only \HI-rich galaxies; no free-floating \HI\ clouds were detected down to 
M$_{HI}\ge$10$^7$\msun.  While these observations are quite sensitive, the data were not obtained nor reduced in a manner conducive
for detecting large-scale, diffuse \HI\ emission.  \citet{hess09} followed up these 
observations by combining 59 hours of archival Australia Telescope Compact Array (ATCA) data with 61 hours of 
GMRT data to produce an extremely sensitive map of NGC~2997 with 5$\sigma$ rms 
sensitivities of 2$\times$10$^5$\msun\ and 9$\times$10$^{18}$\cmsq\ per 6.6\kms\ channel.  
These data reveal the presence of anomalous \HI, some of which is probably related
to ongoing gas accretion.  While it is tempting to associate this accreted
gas with a past interaction, the nearest galaxy is a dwarf $\sim$138 kpc away.
It is possible, however, that these galaxies are connected via a low column 
density \HI\ filament, as is seen between M~31 and M~33.  

NGC~6946 is also a nearby, D$\sim$6 Mpc, late-type, SABcd, spiral galaxy that
resides in loose group \citep{rivers99,karachentsev00}.  It has M$_B$=-21.38 mag,
brighter than M~31, and a rotation velocity of $\sim$160 \kms, corresponding to M$_{dyn}$=9.7$\times$10$^{10}$\msun\ \citep{carignan90}, and a star 
formation rate of $\sim$4 \msun\ yr$^{-1}$ \citep{boomsma08}.  \citet{karachentseva98} and \citet{huchtmeier97}
identified three gas-rich companions within $\sim$50 kpc (projected) from NGC~6946:  UGC~11583, HKK97~L149, HKK97~L150.  \HI\ mapping of NGC~6946 and
companions by \citet{pisano00} using the DRAO synthesis telescope revealed no signatures of interactions in the system.  
\citet{begum04} see a hint of mild warps in both UGC~11583 and HKK97~L149 based on their morphology or kinematics, but there
is no obvious signature of a recent interaction.  \citet{boomsma08} conducted
very sensitive WSRT observations of NGC~6946 revealing many \HI\ clouds
associated with the star-forming disk, and likely originating in a galactic
fountain.  \citet{boomsma08} also found a plume of \HI\ with a similar velocity
to its companions and extending in their general direction, but not actually
connected to those companions down to a column density sensitivity of 
5$\times$10$^{19}$cm$^{-2}$.   Since NGC~6946 is prolifically-forming stars \citep{karachentsev05}
and has had nine supernovae observed in the past century, the \HI\ clouds may be tracing 
star-formation or supernovae-driven outflows, or, alternatively, the star formation may be driven by the inflow
of these \HI\ clouds from the IGM or a past tidal interaction.  The observations I report here will determine if the \HI\ clouds around
NGC~2997 and NGC~6946 are associated with any low column density, diffuse \HI\ structures.  

\section{Observations and Reductions}
\label{sec:obs}

Radio interferometers are powerful instruments delivering excellent resolution and point source sensitivity while providing a map of 
sources across their primary beam.   Because there is a minimum spacing between telescopes in an
interferometer, however, there is a limit to the largest sources they can detect.  Single-dish telescopes, while lacking the resolution of 
interferometers, have much better surface brightness sensitivity and can detect structures on all
angular scales.  Therefore, to search for low column density \HI\ around NGC~2997 and NGC~6946, I used the Robert C. Byrd Green Bank 
Telescope (GBT) with its L-band receiver to map a $2^\circ \times 2^\circ\ $ area centered on both galaxies 
during five observing sessions\footnote{Taken as part of GBT project GBT09B-016.} between 2009 May 9 - June 29.  The GBT is the ideal telescope for
this work with its unique combination of angular resolution (9.1$\arcmin$) and sensitivity (T$_{sys}\sim$20 K); it is the largest, most sensitive radio telescope that can observe these two galaxies.  
At the distance of NGC~2997, the four square degree survey region corresponds to an area of 0.175 Mpc$^2$; at 
the distance of NGC~6946, this is an area of 0.044 Mpc$^2$.  The map was made by 
scanning the telescope along lines of constant right ascension and declination making a ``basket-weave" map of the area of interest.  Each row 
or column was offset by 3\arcmin.  In the direction of the scan, a 5 second integration was 
dumped every 100\arcsec\ (1.67\arcmin), thus assuring that we were Nyquist sampled in both directions.  The GBT spectrometer was used 
with a 12.5 MHz bandwidth, 8192 channels, and 9-level sampling.  The band was centered at the frequency of the \HI\ line at the redshifts of NGC~2997 
(V$_\odot$ = 1088 \kms) or NGC~6946 (V$_\odot$ = 48 \kms).  During the scan, the band was frequency-switched $\pm$2 MHz from this center 
frequency with a one second period for calibration purposes..  These observing techniques allow one to recover large-scale emission from NGC~2997 and NGC~6946 across the entire map.  

For each observing session, I observed either 3C48, 3C147, or 3C295 as a primary flux calibrator using fluxes from \citet{ott94} in order to determine the T$_{cal}$ values 
for the noise diode.  The resultant values were constant between sessions with T$_{cal}$= 1.53 K and 1.54 K for the XX and YY polarizations with an error of 1\%.  
An aperture efficiency of $\sim$0.66, appropriate for the GBT at 1420 MHz, was used for the calibrations \citep{boothroyd11} leading to a gain of 2 K Jy$^{-1}$.  The typical system temperature for the 
observations was $\sim$20 K.  The frequency-switched spectra were reduced in the standard manner using the {\it getfs} procedure in the GBTIDL\footnote{http://gbtidl.nrao.edu/} 
data reduction package.  Because of its redshift, for NGC~2997 the frequency-switched data were not folded, since folding them placed the negative 
Galactic \HI\ emission on top of NGC~2997 itself.  This was not a problem for NGC~6946, so the data were folded producing improved noise by a factor
of $\sqrt{2}$.  A third order polynomial was fit to the line-free regions of the spectra to remove any residual baseline structure and continuum sources.  
I assumed a constant zenith opacity of 0.01 \citep[appropriate at 21~cm, e.g.][]{chynoweth08} to convert the calibrated data into units of T$_A^*$.  About 0.4\% of all the integrations were flagged due to 
broadband RFI.  Calibrated data were boxcar smoothed to a velocity resolution of $\sim$5.15 \kms, and a velocity range of 900 to 1300 \kms\ 
and -300 \kms\ to 300 \kms\ was exported from GBTIDL for NGC~2997 and NGC~6946, respectively.  The calibrated data were converted into an 
appropriate format for gridding using the {\sc idlToSdfits}\footnote{developed by Glen Langston of NRAO; documentation at http://wiki.gb.nrao.edu/bin.view/Data/IdlToSdifts.} 
program and then imported into AIPS where it was gridded into a map using the task SDGRD with a convolution function of a Gaussian-tapered circular Bessel function \citep{mangum07}.  
For NGC~2997, an additional fourth order baseline was removed from the data cube using the AIPS task XBASL; no additional 
baseline removal was needed for the NGC~6946 data.  To facilitate comparison with the previous WSRT observations of NGC~6946, the data cube was converted to units of mJy/beam and 
resampled using the Miriad \citep{sault95} task, REGRID, to a channel spacing of 4.2 \kms.

For the final map of NGC~2997, the rms noise is 21 mK per 5.15 \kms\ channel equivalent to \nhi = 2.0$\times$10$^{17}$\cmsq\ for 
optically-thin emission filling the 9.2\arcmin\ GBT beam, or a 5$\sigma$ detection limit for an unresolved source with a linewidth of 20~\kms\ of \mhi = 2$\times$10$^7$\msun.  The \nhi\ 
sensitivity is 45$\times$ better than the \citet{hess09} survey, while the \mhi\ sensitivity is 100$\times$ worse.  The final map of NGC~6946 has a rms noise of 15 mK per 5.15 \kms\ channel 
equivalent to \nhi = 1.4$\times$10$^{17}$\cmsq\ for optically-thin emission filling the GBT beam or a 5$\sigma$ detection limit (as described above) of \mhi = 3$\times$10$^6$\msun.

To confirm the reality of an apparent \HI\ filament associated with NGC~6946 an additional pointed observation towards $\alpha (J2000) =$ 20:31:56.5, $\delta (J2000) =$ 60:21:27 
was made on 2009 July 5.  The observation was position-switched:  observing the plume for 5 minutes followed by blank sky offset by 1.2\arcdeg\ in right 
ascension for 5 minutes.  This cycle was repeated for $\sim$2 hours.  The {\it getps} procedure in GBTIDL was used to reduce the observations using the 
same T$_{cal}$, aperture efficiency, and opacity values listed above.  After averaging both polarizations and boxcar-smoothing the final spectrum to 
5.15 \kms\ channels, the final rms noise was 3.6 mK (\nhi = 3.3$\times$10$^{16}$\cmsq) per 5.15 \kms\ channel.  

\section{Results}
\label{sec:res}

\subsection{NGC~2997}

\begin{figure}
\includegraphics[width=0.9\textwidth]{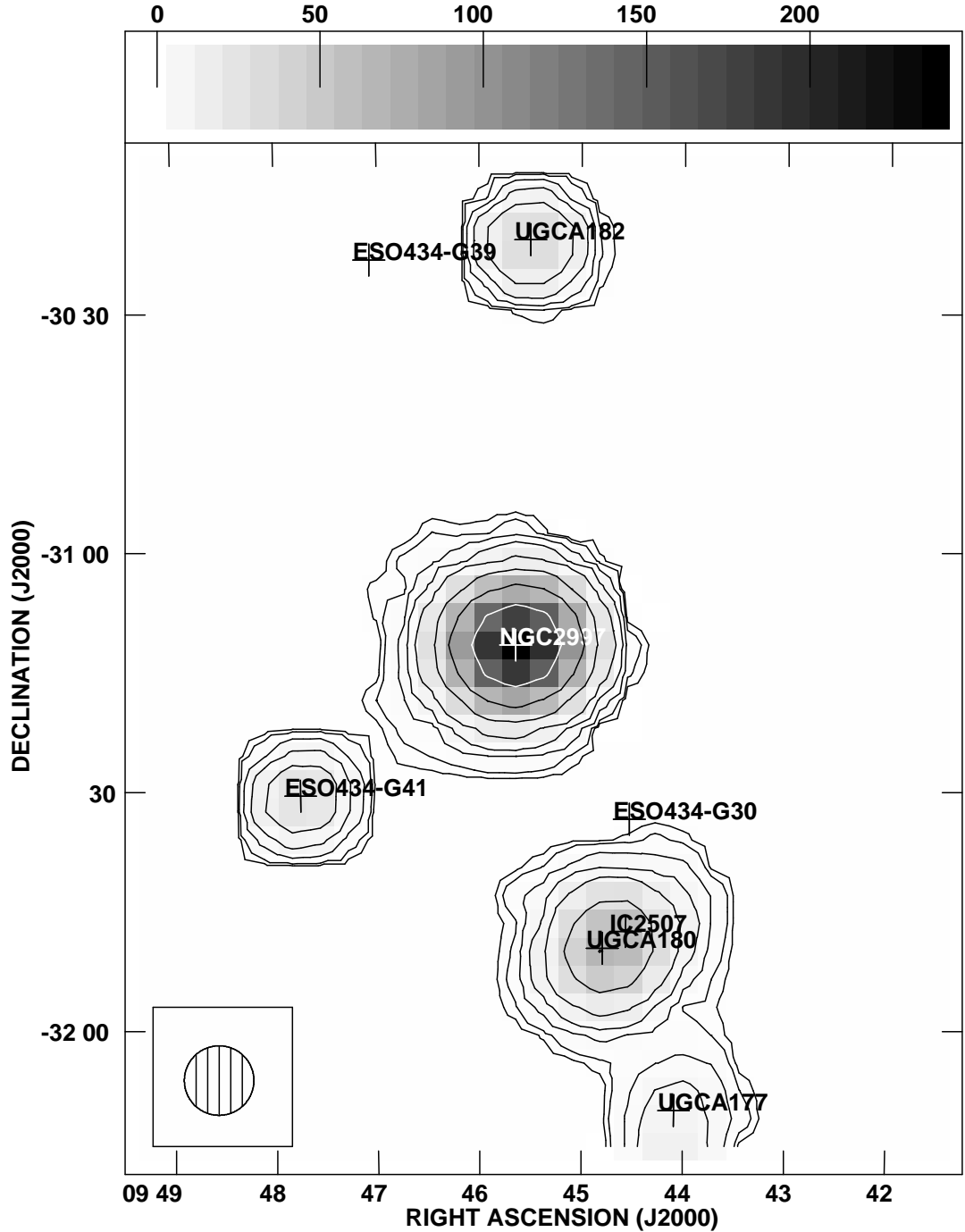}
\caption{The GBT total \HI\ intensity (moment 0) map of the four square degree region around NGC~2997.  The greyscale is in units of K \kms, which is 
proportional to \nhi.  Contours start at 1.2$\times$10$^{18}$\cmsq\ (equivalent to a 3$\sigma$ detection for a 20 \kms\ width line) and continue at 
2, 5, 10, 20, 50, 100, 200, 500, 1000, and 2000 times that level.  Optically-identified group members are marked with plus signs. 
There are no visible filamentary structures connecting the group galaxies; the apparent connection between UGCA~177 with UGCA~180 and IC~2507 is
not contiguous in velocity.  \label{fig:ngc2997}}
\end{figure}

Figure~\ref{fig:ngc2997} shows the total \HI\ intensity map from the GBT observations of NGC~2997 and its surroundings.  The observations reveal all of 
the galaxies detected by \citet{pisano11} using Parkes and the ATCA, but no additional emission.  The only known group galaxies in the field that were 
undetected are ESO~434-G39 at V$_\odot$=1024 \kms, and ESO~434-G30 at V$_\odot$ = 1288 \kms.  The GBT observations show no 
clear connections between discrete galaxies; there is no \HI\ filament connecting NGC~2997 with any of its neighbors down to a 5$\sigma$, 20 \kms\ detection limit 
of \nhi =1.2$\times$10$^{18}$\cmsq.  The blending of the \HI\ contours between UGCA~180, IC~2507 and UGCA~177 is not evident in the individual 
channel maps and appears to just be extended \HI\ emission associated with the individual galaxies.  

\begin{figure}
\includegraphics[width=0.9\textwidth]{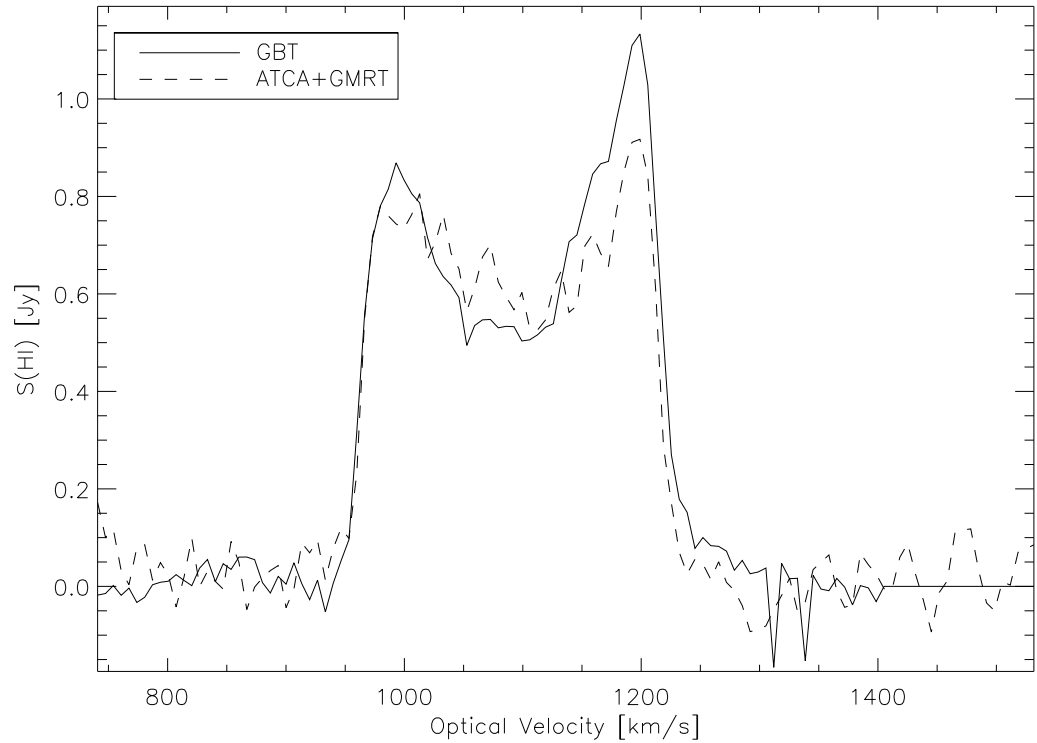}
\caption{A comparison the HI spectrum of NGC 2997 from the GBT map (solid line) and from the GMRT+ATCA \citet{hess09} data (dashed line).  Both spectra are integrated over
the same area (0.24 square degrees) and have been sampled on with the same channel spacing.  It is clear that the two spectra agree within the noise except around 1200 \kms\
where the GBT has recovered more flux.  \label{fig:ngc2997_spec}}
\end{figure}

While no clear filamentary connections are seen between the galaxies, it is possible that there is an extended reservoir of \HI\ around the individual galaxies.  
there are two ways to check for this:  one can compare the \HI\ areal coverage or the integrated \HI\ fluxes for individual galaxies.  High quality, sensitive data only 
exists for NGC~2997, so this analysis will be limited to this galaxy; this comparison cannot be done for the other galaxies in the field since the ATCA observations 
of \citet{pisano11} are much noisier.  The integrated \HI\ flux of NGC~2997 from the combined GMRT and ATCA data at low
resolution (188$\arcsec \times$110$\arcsec$) presented by \citet{hess09} yields S$_{HI}=$180.52$\pm$0.21 Jy \kms.  For the same area, the GBT data yields
S$_{HI}=$194.52$\pm$0.18 Jy \kms (not including the 1\% error in the flux calibration), which is 7.2\% higher than the interferometer flux.  The two spectra are shown in Figure~\ref{fig:ngc2997_spec}, clearly showing that
they are in general agreement except around 1200 \kms\, where the GBT \HI\ flux is clearly higher.  This velocity corresponds to that of the east side of NGC~2997, 
possibly corresponding to the lopsided \HI\ distribution at low column density seen in Figure~\ref{fig:ngc2997}.  

As for the areal extent of the \HI\ distribution from the two datasets, we placed both datasets on the same pixel and velocity scale, as above, and convolved the
\citet{hess09} data to match the resolution of our GBT data.  While this improves the column density sensitivity of the \citet{hess09} cube, those data are still insensitive
to emission on angular scales greater than $\sim$20$^\prime$.  To measure the areal extent of both datasets, we then created a total \HI\ intensity (moment 0) map 
blanked at the 2$\sigma$ level based on the noise measured in each cube.  The \nhi\ limit is set by the \citet{hess09} data and is 1.2$\times$10$^{18}$\cmsq\ at the
3$\sigma$ level per 6.6 \kms\ channel.  The areas of both maps above this \nhi\ level, and their ratios, are listed in Table~\ref{tab:areas}.  In general, the area of NGC~2997 
as measured with the GBT at a given \nhi\ is larger than that from the interferometer data.  The only exceptions are at 10$^{19}$\cmsq\ and 5$\times$10$^{19}$\cmsq.  These
anomalies are due to the presences of artifacts in the interferometer data that dominate the area of emission at these \nhi\ levels, and is not a physical effect.  

We can also examine how the \HI\ extent of NGC~2997 grows as compared to theoretical predictions by \citet[][see their Figure 6]{popping09}.  Normalizing to the area of
NGC~2997 at \nhi $=$10$^{20}$\cmsq, the predicted cumulative area (listed in Table~\ref{tab:areas}) is smaller than the measured area by $\sim$20-40\% at all lower column densities \citep{popping09}.
Given the uncertainty of the assumptions used for the simulations, this indicates reasonably good agreement.  This is true whether we use the interferometer data or the GBT data alone for the comparison.  

\subsection{NGC~6946}

\begin{figure}
\includegraphics[width=0.9\textwidth,angle=-90]{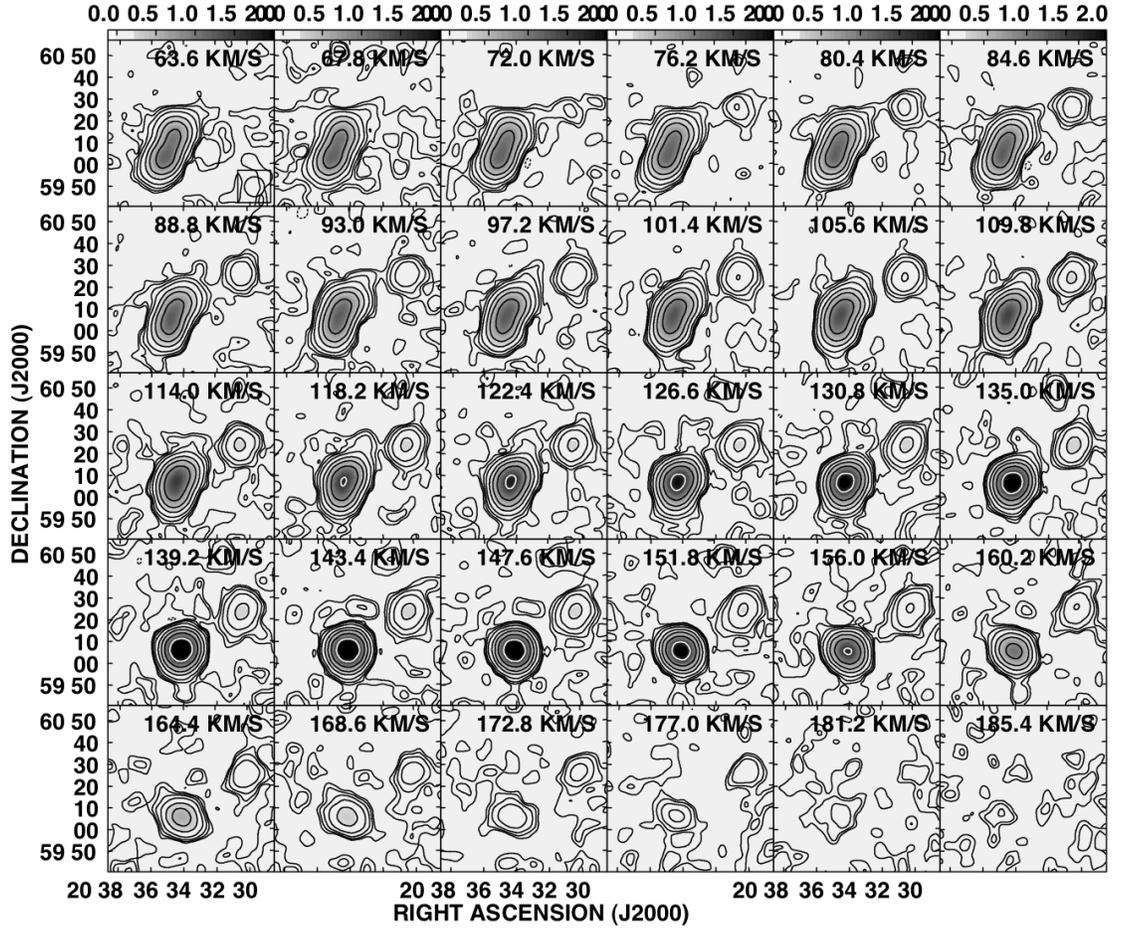}
\caption{GBT \HI\ channel maps of NGC~6946, its companions, and \HI\ filament.  Contours are at -2, 2, 3, 5, 10, 25, 50, 100, 200$\sigma$, where
$\sigma$=8 mJy/beam over a 4.2 \kms\ channel.   The greyscale has units of mJy/GBT beam.  \label{fig:ngc6946chan}}
\end{figure}

\begin{figure}
\includegraphics[width=0.9\textwidth]{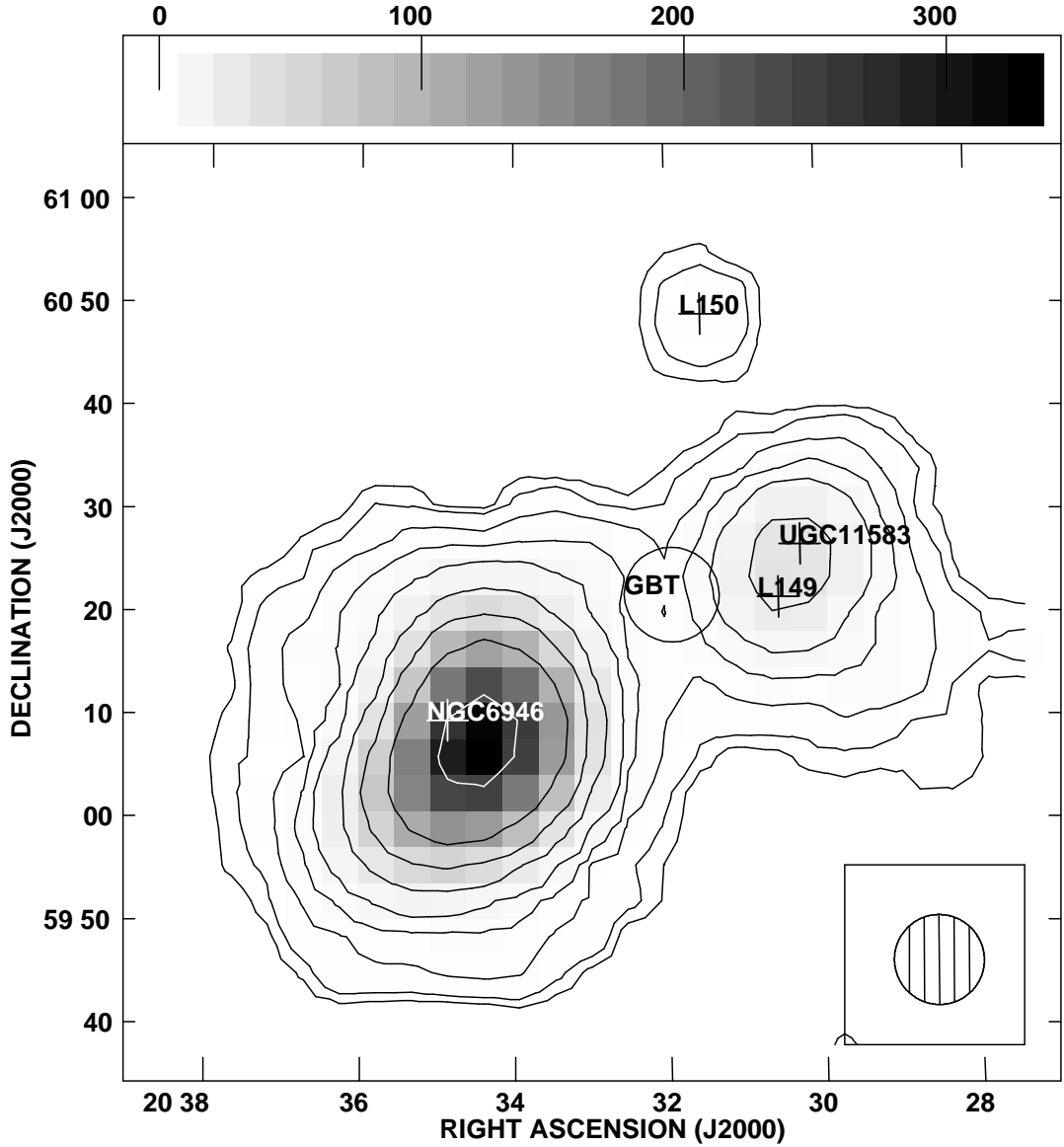}
\caption{GBT total \HI\ intensity map of NGC~6946, its companions, and putative \HI\ filament. Contours are at 0.1, 0.2, 0.5, 1, 2, 5, 10, 20, 500$\times$10$^{19}$\cmsq.  
Greyscale is in units of K \kms, proportional to the \HI\ column density.  The GBT beam is shown in the lower right.  The known galaxies are marked with
plus signs.  The GBT label indicates the position for the spectrum shown in Figure~\ref{fig:ngc6946spec}.  \label{fig:ngc6946m0}}
\end{figure}

\begin{figure}
\includegraphics[width=0.9\textwidth]{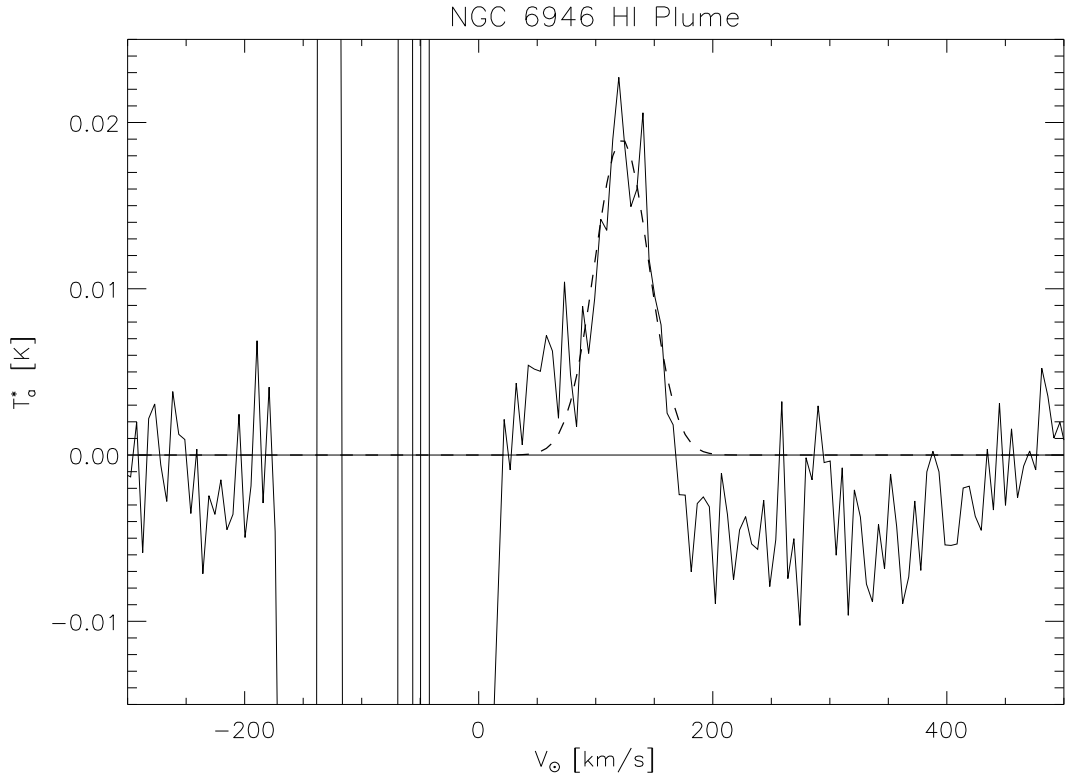}
\caption{GBT spectrum of  the \HI\ filament. The dashed line indicates the best fitting single-component Gaussian fit to the line with a peak of 19$\pm$2 mK, a
center of 122.2$\pm$2.4 \kms, and a FWHM linewidth of 54.5$\pm$5.7 \kms.  \HI\ emission associated with the Milky Way spans V$_\odot$= -170 -- 50 \kms.  \label{fig:ngc6946spec}}
\end{figure}

Channel maps of NGC~6946 and its surroundings are shown in Figure~\ref{fig:ngc6946chan} down to 63.6 \kms\ after which emission from NGC~6946 begins to be confused
with the Milky Way.  These maps show emission from NGC~6946 and three of its companions: L150 and L149 \citep[first discovered by][]{huchtmeier97} and UGC~11583; the latter
two of which are confused within the GBT beam.  The channel maps also reveal an \HI\ filament, present at the 2-3$\sigma$ level in each of $\sim$12 channels, connecting NGC~6946 
with UGC~11583 and L149.  This feature is more clearly seen in the total \HI\ intensity map in Figure~\ref{fig:ngc6946m0}, and is completely invisible in the WSRT data from
\citet{boomsma08} down to \nhi$\sim$5$\times$10$^{19}$\cmsq.  From Figure~\ref{fig:ngc6946m0}, this filament has a peak \nhi$\sim$5$\times$10$^{18}$\cmsq\ as measured with the GBT at a resolution 
of 9.2$\arcmin$ (16 kpc).  At this level, the filament
is barely resolved with a width of 22 kpc (12.5$\arcmin$) and a length of $<$12 kpc (6.8$\arcmin$), uncorrected for the GBT beamsize.  If our observations of this feature suffer from beam dilution, then the 
filament would be smaller, but would have a \nhi\ larger by the ratio of the GBT beam area to the true source area.  As such, it cannot have a width smaller than $\sim$2 kpc (1$\arcmin$) or
its column density would have been high enough ($\gtrsim$10$^{20}$\cmsq) to be detected by \citep{boomsma08}.  To confirm the reality of this filament, I conducted a pointed observations 
with the GBT at the position marked in Figure~\ref{fig:ngc6946m0}.  The spectrum from the pointed observation of the \HI\ filament connecting NGC~6946 with UGC 11583 and L149 is 
shown in Figure~\ref{fig:ngc6946spec}.  The large wiggles between $\sim$-200 \kms\ and 30 \kms\ are from the Milky Way \HI\  that is not completely subtracted by position-switching.  
The \HI\ filament is clearly visible as an emission line at 123$\pm$2 \kms, clearly distinct from Milky Way \HI\ emission.  A Gaussian fit to this line shows that it has a 
FWHM = 55$\pm$5 \kms, a peak at 18$\pm$1.6 mK, and an integrated column density of 0.92 K \kms\ or 1.7$\times$10$^{18}$\cmsq.  It is clear from these data that the filament is 
not just structure in the baseline, but is real emission.  

\begin{figure}
\includegraphics[width=0.9\textwidth,angle=-90]{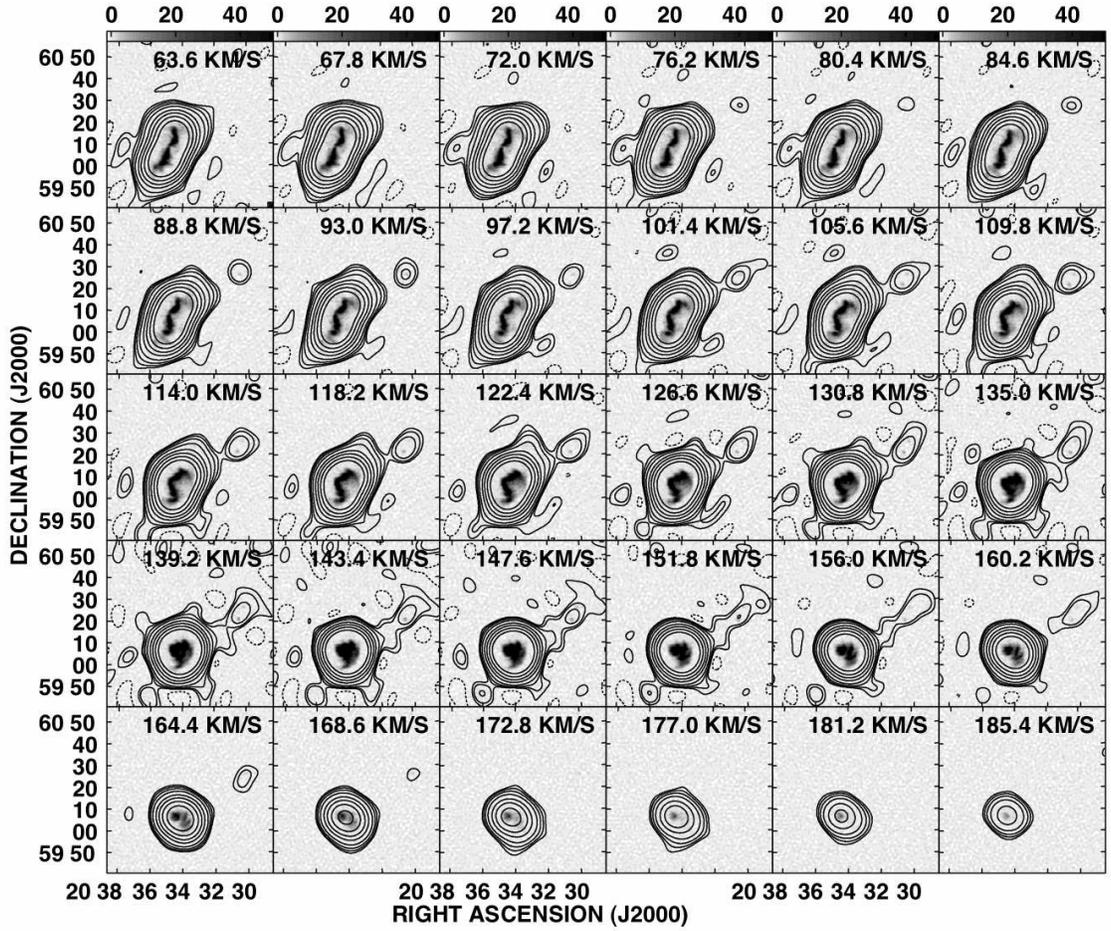}
\caption{WSRT \HI\ channel maps of NGC~6946, its companions, and \HI\ filament from \citet{boomsma08}.  The greyscale represents the WSRT data with units of mJy/beam at 64$\arcsec$ resolution 
on a logarithmic scale.  The contours show the WSRT data when convolved to the GBT beamsize and are at -2, 2, 3, 5, 10, 25, 50, 100, 200$\sigma$, where $\sigma$= 1 mJy/GBT beam.  A filamentary feature is 
clearly visible, but it is about an order of magnitude fainter than seen in the GBT data.  \label{fig:ngc6946wsrt}}
\end{figure}

\begin{figure}
\includegraphics[width=0.9\textwidth,angle=-90]{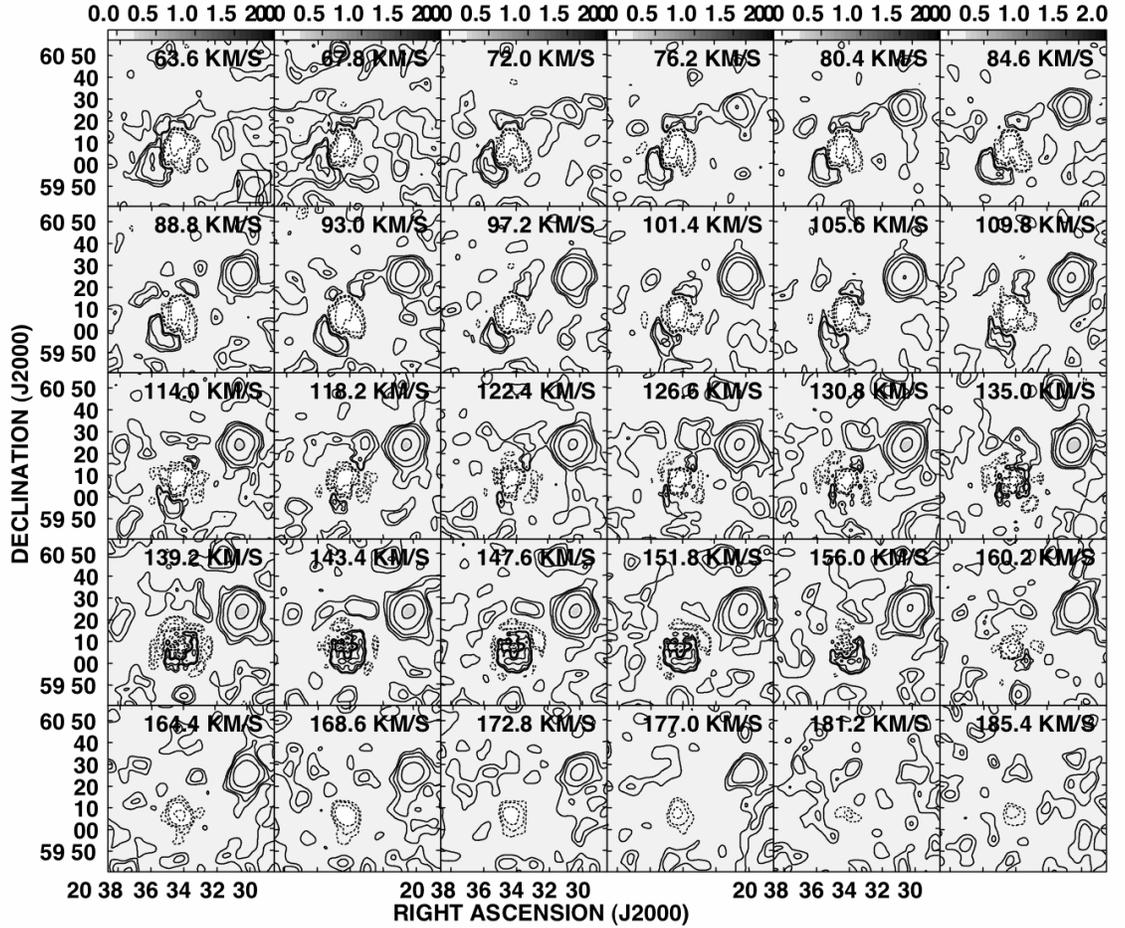}
\caption{Channel maps showing the convolved \HI\ data subtracted from the GBT \HI\ data for NGC~6946, its companions, and \HI\ filament.  Contours are at -10,-5,-3,-2, 2, 3, 5, 10, 25, 50, 100, 200$\sigma$, where
$\sigma$=8 mJy/beam.   The greyscale has units of mJy/GBT beam.  \label{fig:ngc6946diff}}
\end{figure}

This putative filament is also not the product of stray radiation.  The near sidelobes of the GBT are less than one percent of the main beam 
response and are symmetric \citep{boothroyd11}, and since there are no other observed features at this level around NGC~6946 or its companions it is unlikely that this feature 
is due to stray radiation.  To confirm this, I convolved the WSRT data of \citet{boomsma08} with a Gaussian GBT beam.  A filamentary feature connecting NGC~6946 and its companions appears, as seen in 
Figure~\ref{fig:ngc6946wsrt}, but at a \nhi\ level an order of magnitude lower than the observed \nhi\ of this \HI\ filament.  The \HI\ filament becomes even more visible when subtracting the convolved 
WSRT \HI\ data from the GBT \HI\ data as shown in Figure~\ref{fig:ngc6946diff}.  Unfortunately, due to the relatively small extent of this filamentary feature compared to the GBT 
beamsize, it is impossible to know for sure if this filament actually connects NGC~6946 with its companions or is merely a low \nhi\ extension of the \HI\ plume seen by \citet{boomsma08}.

Since NGC~6946 is confused with Milky Way \HI\ emission below $\sim$50 \kms, it is not feasible to compare the extent of the \HI\ in the WSRT and GBT observations.  
Nor is it feasible to compare the integrated \HI\ flux as observed by the two telescopes as was done for NGC~2997.  

\section{Discussion}
\label{sec:disc}

These GBT observations of NGC~2997 and NGC~6946 reveal an extended low-\nhi\ distribution around the former galaxy and a filamentary structure attached to 
the latter.  What is the nature of these extended \HI\ structures around these two galaxies?  In the case of NGC~2997, the low \nhi\ gas is morphologically and 
kinematically similar to the galaxy's disk.  The slight lopsidedness may due to a past interaction \citep[but see][]{zaritsky13}, the unresolved, high column density 
peak of a filamentary cold flow, tidal debris, or just an extended \HI\ disk.  NGC~2997 has a dynamical mass, M$_{dyn}$= 2.1$\times$10$^{11}$\msun\ \citep{hess09}, 
that is low enough to expect that cold flows are its dominant mode of accretion \citep[e.g.][]{keres05}.  NGC~2997 is located within a loose galaxy group, LGG~180, with an estimated 
M$_{virial}\sim$7$\times$10$^{12}$\msun\ \citep{pisano11} so that it may reside in a sub-halo within a more massive group halo that suppresses cold accretion.  In addition, the group galaxies 
have a mean separation of $\sim$500 kpc so tidal interactions with neighboring galaxies are relatively rare, but certainly possible \citep{pisano11}. The general agreement 
in the distribution of \HI\ column density with the predictions of \citet{popping09} suggest that this is not a unique feature so an extended \HI\ disk is a real possibility.  The absence 
of any filamentary features may be due to the physical reasons listed above or, since NGC~2997 is more distant than NGC~6946 and the GBT beamwidth is $\sim$32 
kpc at D$\sim$12 Mpc, possibly the result of beam dilution; deeper observations at higher resolution are needed to test these possibilities.  The slight excess of \mhi\ in the
GBT data relative to the interferometer data reflects the fact that there is very little \HI\ present around NGC~2997.  This is not surprising as at these low column densities, most of the
hydrogen present will be ionized.  In fact, we know from absorption line studies that the covering fraction of \HI\ at $log$\nhi$\gtrsim$13 is 100\% within $\sim$200 kpc  \citep{tumlinson13}.  
As such, the \HI\ emission at low \nhi\ serves as a tracer of the bulk of the ionized gas, which may represent over 70\% of the total hydrogen mass around galaxies \citep{popping09}.  

The \HI\ filament between NGC~6946 and its companions could be a tidal bridge or a cold flow that is part of the cosmic web.  NGC~6946 has a dynamical mass even lower than 
NGC~2997, so cold mode accretion is certainly feasible.  Furthermore, while NGC~6946 is in a loose group, this group is less massive than LGG~180 with M$_{virial}\sim$1.6$\times$10$^{12}$\msun, although the galaxy
density is about 10 times higher, making a tidal origin plausible \citep{rivers99,karachentsev00}.  Dividing the width of the filament, 22 kpc, by its linewidth, 55 \kms, yields a dispersion timescale of $\sim$400 Myr.  
This width may be too large, however, if the filament is beam-diluted then it could be as narrow as 2 kpc.  In this case, the dispersion timescale would only be $\sim$40 Myr.  While the latter timescale is very short, the former is 
comparable to the interaction timescale: the projected on-sky separation of the nearest companions divided by their line-of-sight velocity separation from NGC~6946, $\sim$700-800 Myr; particularly accounting for the
unknown real, physical separations and velocity differences.  The best way to confirm that this \HI\ filament is due to a past encounter between NGC~6946 and its satellites would be to identify an associated stellar stream 
\citep[e.g.][]{ibata01,ibata07,ferguson02}.  Unfortunately, due to the low Galactic latitude of NGC~6946, 11.6$\arcdeg$, and high extinction, A$_V$=0.94 mag, identifying such streams will be challenging. 

Finally, it is worth noting that both NGC~2997 and NGC~6946 have extensive populations of high velocity clouds, many of which are coincident with the stellar disk and, hence, are likely related to a galactic fountain, 
but some of which are consistent with ongoing gas accretion \citep{hess09, boomsma08}.  Aside from the \HI\ plume identified by \citet{boomsma08} in the outskirts of NGC~6946, these features have no relationship 
to the presence or absence of diffuse \HI\ filaments seen with GBT.  This suggests that the diffuse \HI\ features and the bulk of the high velocity clouds seen by the interferometer observations have different origins.  

\section{Conclusions}
\label{sec:conc}

In this paper, I reported the results from a pilot GBT \HI\ survey of a four square degree region surrounding the nearby galaxies NGC~2997 and NGC~6946 down to a detection limit of \nhi$\sim$10$^{18}$\cmsq\ over a linewidth
of 20 \kms.  The goal of this survey was to identify \HI\ filaments associated with predicted cold mode accretion and analogous to the \HI\ features seen around M~31.  Such features are large on the sky, and, therefore, 
invisible to radio interferometers due to the gaps in the {\it{uv}}-coverage of all interferometers.  In contrast, single-dish telescopes like the GBT are sensitive to emission on all spatial scales and have excellent surface
brightness sensitivity.  The observations of NGC~2997 reveal a more extended \HI\ distribution around the galaxy, but no filamentary features.  This extended \HI\ may be associated with an extended \HI\ disk, a past 
tidal interaction or the highest column density parts of a cold, filamentary flow.  Observations of NGC~6946 reveal a \HI\ filament that appears to connect it with its neighbors.  Morphologically this is similar to a predicted 
cold flow, but the similarity of the filament's dispersion timescale with the interaction timescale suggest that it has a tidal origin.  While detecting an associated stellar stream would strongly suggest that this \HI\ filament 
is tidal in origin, the requisite observations do not yet exist.  

These observations illustrate the biggest problem in trying to identify cold flows via \HI\ emission, that is there are many plausible origins for features with similar properties.  The best
way to identify cold flows is statistically via a large survey.  Cold flows should preferentially exist around low mass galaxies in low density environments.  Tidal debris, on the other hand, should exist preferentially around
galaxies in high density environments.  Galactic outflows should preferentially exist around galaxies with low masses and large star formation rates.  The ideal \HI\ survey would span a wide range of galaxy mass, environment,
and star formation rates to discriminate between these different origins.  Such a survey is best done with single-dish radio telescopes that are
large (so that they have good angular resolution) with low sidelobes (so that there is no confusion with \HI\ emission from the galaxy itself) and are extremely sensitive (so that low column density features can be easily
detected).  The GBT remains the only telescope in the world that combines all of these features.  Pisano et al. (2014, in preparation) will report on the results from just such a GBT survey of the THINGS galaxies
\citep{walter08}.

\acknowledgements

I wish to acknowledge the excellent staff at the Green Bank Telescope for their work in keeping the telescope operating at a high level of performance and making it 
so easy to use.  I would also like to thank them for their prompt scheduling of these observations and generously providing additional observing time for confirmation 
observations.  Thanks to Jay Lockman, Glen Langston, and Katie Keating for their advice on reducing GBT mapping data and to all the Green Bank scientific staff 
for helpful suggestions for analysis of the data.  Thanks also to Tom Oosterloo for providing the WSRT data cube for comparison with the GBT data, and to Frank Briggs, Filippo Fraternali, and Tom Oosterloo for
helpful conversations about these data.  Thanks to David Frayer and Jay Lockman for their helpful comments on this paper, and to the anonymous referee for his/her constructive report, which helped improve the paper.  
This work was partially supported by NSF CAREER grant AST-1149491.  
The National Radio Astronomy Observatory is a facility of the National Science Foundation operated under cooperative agreement by Associated Universities, Inc.

{\it Facility:} \facility{GBT}

\clearpage

\begin{deluxetable}{ccccccc}
\tablecolumns{7}
\tablewidth{0pc}
\tablecaption{Cumulative Area of \HI\ for NGC~2997 \label{tab:areas}}
\tablehead{\colhead{\nhi} & \multicolumn{2}{c}{GBT area} & \multicolumn{2}{c}{GMRT+ATCA area} & \colhead{$\frac{GBT}{GMRT+ATCA}$ area} & \colhead{Predicted Cum. Area\tablenotemark{a}} \\
\colhead{\cmsq} & \colhead{$\square^\prime$} & \colhead{Normalized} & \colhead{$\square^\prime$} & \colhead{Normalized} & \colhead{} & \colhead{}}
\startdata
$\ge$1.2$\times$10$^{18}$ & 704 & 3.1 & 573 & 2.7 & 1.23 & 2.40 \\
$\ge$5$\times$10$^{18}$ & 588 & 2.6 & 547 & 2.5 & 1.07 & 1.80 \\
$\ge$1$\times$10$^{19}$ & 511 & 2.2 & 520 & 2.4 & 0.983 & 1.58 \\
$\ge$5$\times$10$^{19}$ & 324 & 1.4 & 343 & 1.6 & 0.945 & 1.19 \\
$\ge$1$\times$10$^{20}$ & 229 & 1.0 & 215 & 1.0 & 1.07 & 1.00 \\
\enddata
\tablenotetext{a}{From \citet{popping09}.  The cumulative area is normalized to 1.00 at \nhi$=$10$^{20}$\cmsq.}  
\end{deluxetable}

\clearpage

\end{document}